\documentclass[aps,prb,prbbib,twocolumn,epsf]{revtex4}
\usepackage{graphicx}
\usepackage{bm}
\usepackage{setspace}
\begin{document}
\title{Current-induced torques and interfacial spin-orbit coupling}
\author{Paul M. Haney$,^{1}$ Hyun-Woo Lee,$^{2}$ Kyung-Jin Lee,$^{1,3,4,5}$ Aur\'{e}lien Manchon,$^{6}$  and  M. D. Stiles$^{1}$}
\affiliation{
$^1$Center for Nanoscale Science and Technology, National
Institute of Standards and Technology, Gaithersburg, Maryland 20899-6202, USA \\
$^2$PCTP and Department of Physics, Pohang University of Science and Technology, Kyungbuk 790-784, Korea\\
$^3$Korea University, Department of Material Science \& Engineering, Seoul 136701, South Korea\\
$^4$Korea Institute of Science and Technology, Seoul 136-791, Korea\\
$^5$Univeristy of Maryland, Maryland Nanocenter, College Pk, MD 20742 USA\\
$^6$Core Labs, King Abdullah University of Science and Technology (KAUST), Thuwal 23955-6900, Saudi Arabia
 }

\begin{abstract}
  In bilayer systems consisting of an ultrathin ferromagnetic layer
  adjacent to a metal with strong spin-orbit coupling, an applied
  in-plane current induces torques on the magnetization.  The torques
  that arise from spin-orbit coupling are of particular interest.
  Here, we calculate the current-induced torque in a Pt-Co bilayer
  to help determine the underlying mechanism using first principles methods.  We focus exclusively on the analogue to the Rashba torque, and do not consider the spin Hall effect.  The details of the torque depend strongly on the layer thicknesses
  and the interface structure, providing an explanation for the wide
  variation in results found by different groups.  The torque depends
  on the magnetization direction in a way similar to that found for a
  simple Rashba model.  Artificially turning off the exchange
  spin splitting and separately the spin-orbit coupling potential in
  the Pt shows that the primary source of the ``field-like" torque is
  a proximate spin-orbit effect on the Co layer induced by the strong
  spin-orbit coupling in the Pt.
\end{abstract}

\pacs{
85.35.-p,               %nanoelectronic devices
72.25.-b,               %spin polarized transport
} \maketitle

\section{Introduction}

Spintronics has had significant impact on information technology, the
most common example being the read heads in hard disk drives.  The field is
poised to have even greater impact, led by the development of devices
such as spin transfer torque-based magnetic random access memory.  A
crucial component of this next generation of spintronic applications
is the control of magnetic orientation with an electrical
current\cite{stiles,katine} (or electric field\cite{chien,ohno0}).  The
approach is furthest developed in magnetic tunnel junctions, which
utilize spin transfer torque to reversibly switch the magnetization in
one of the layers.  An alternative approach has been demonstrated in
recent experiments on bilayer systems consisting of ultrathin
ferromagnetic layers adjacent to heavy metals such as
Pt.\cite{miron1,miron2,miron3,liu1,ando,kim0,parkin,beach,fan}  In these systems, spin-orbit
coupling is responsible for current-induced torques.  There are
indications that the efficiency of these current-induced torques
(measured as, for example, the torque per current density) may be
larger than the conventional spin transfer torque.\cite{ralph}  For
this reason among others, these bilayer and related systems offer a
possible route for spintronics to fulfill its full promise in
technological applications.

The spin-orbit coupling in these systems leads to multiple effects.
For one, there is a spin Hall effect in the Pt layer, so that spin
current flows from the Pt into the magnetic layer, with spin
orientation perpendicular to the charge current direction and the
interface normal.  This spin current flux induces magnetization
dynamics via the conventional spin transfer torque.  A distinct effect
originates from the simultaneous presence of magnetization, spin-orbit
coupling, and broken inversion symmetry at the interface.  These
ingredients result in an electronic structure in which current
carrying states acquire a spin accumulation transverse to the
magnetization, resulting in a torque.\cite{manchon}  Both the spin
Hall effect and interfacial spin-orbit torque are proportional to the
charge current.  An additional consequence of the interfacial spin-orbit
coupling is the current-independent Dzyaloshinskii-Moriya (DM)
interaction,\cite{Dzyaloshinskii,Moriya} which has been argued to be
important\cite{thiaville,parkin,beach} for current-induced domain wall motion.

The vector components of the spin-orbit torques can be
decomposed into two vector fields as a function of ${\bf {\hat M}}$, the
magnetization direction:
$\hat{\bf M} \times \left(\hat{\bf j} \times \hat{\bf z}\right)$, which we refer to as a
field-like torque and
$\hat{\bf M} \times \left[\hat{\bf M} \times \left(\hat{\bf j} \times \hat{\bf
    z}\right)\right]$, which we refer to as a damping-like torque
(${\bf \hat{j}}$ is the charge current direction and ${\hat{\bf z}}$ is the interface
normal).\cite{footnote1}
These names derive from the similarity of the first form to the torque due
to a field along $\hat{\bf j} \times \hat{\bf z}$, and the second to the damping
that would result from precession around that field.

Depending on the assumed scattering processes, both the mechanism that
combines the spin
Hall effect plus spin-transfer torque and the interfacial mechanism
can give torques in both directions.\cite{macdonald,wang,duine2012,Kim2012}  However in
the clean limit of the relaxation time approximation, Freimuth {\it et
  al.}\cite{Freimuth} have shown that the physics behind two torque
components separate rather cleanly:  The damping torque originates
predominantly from the spin Hall effect, and is a consequence of the
perturbation of electronic states by the applied electric field, while
the field-like torque originates mostly from the spin-orbit coupling
at the interface, in conjunction with the perturbation of the electron
distribution function by the applied field.  In this paper, we focus
on the underlying physics behind the interfacial spin-orbit torque,
neglecting the spin Hall effect.  In the terminology of
Ref.~\onlinecite{Freimuth}, we calculate the odd torque only
[$\bm\Gamma(\hat{\bf M})=-\bm\Gamma(-\hat{\bf M})$], and show
that this component yields predominantly ``field-like" torques for realistic systems
(within the relaxation time approximation).

Which of these mechanisms (field-like or damping-like) is responsible for these current-induced
torques is controversial.  Measurements of the reversal of magnetic
layers are interpreted in terms of a damping-like torque due to the
spin Hall effect.\cite{ralph} Some experiments on current-induced
domain wall motion in these systems are interpreted in terms of a
damping-like torque in combination with the DM
interaction.\cite{parkin,beach} Other experiments are interpreted in
terms of a combination of field-like and damping-like torques arising
from the interface.\cite{miron1} Experimental measurements of the
torque vector show varied results for the magnitude of the field-like
and damping torques, depending sensitively on the sample
details.\cite{kim0} Calculations\cite{Freimuth} show both torques to
be present and sensitive to structural details.  The spin-orbit
coupling affects the magnetization dynamics in multiple ways, and
clearly distinguishing the different contributions requires more
careful experimental and theoretical efforts.  In this work, we
calculate the current-induced torques present at the interface between
Co and Pt using first principles methods.  As discussed in
Sec. II, we do not include the contributions from the spin Hall
effect.  The calculation is therefore analogous to that of the 2DEG
Rashba model, but with the full electronic structure of the Co-Pt
interface taken into account.  With this approach, we explore the
sensitivity of the current-induced torques to the system structure,
the angular dependence of the current-induced torques, and attempt to
identify the most important physical ingredients of the system which
lead to the current-induced torques.

\section{Method}

A quantitative description of any system using density functional
theory requires specific knowledge of the atomic structure.  In the
absence of experimental characterization at this level of detail, we study a
variety of structures to reach qualitative conclusions.  Our aim
with this approach is to extract semi-quantitative estimates of the
current-induced torques, and to identify the trends and most important
physical mechanisms underlying the current-induced torques.

\begin{figure}
\includegraphics[width=1\columnwidth]{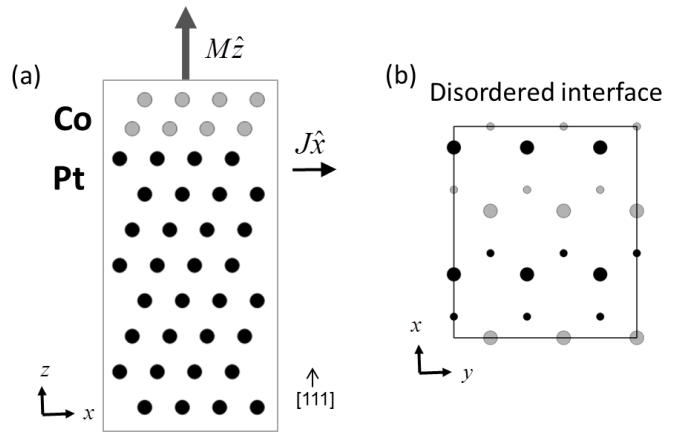}
\caption{(a) System geometry for an ideal interface, (b) Example of an alloyed interface.  Black (gray) dots represent Pt (Co), and the larger dots are in a layer above the smaller dots.}\label{fig:coords}
\end{figure}

There is significant mismatch between the lattice constants of bulk Co
and Pt ($a_{\rm Co}=0.354~{\rm nm},~a_{\rm Pt}=0.392~{\rm nm}$).
Studies of Co growth on Pt observe a structure that is generally
inhomogeneous and quite sensitive to the Co coverage.\cite{ferrer,lundgren} For
simplicity, we assume a uniform Co layer with the in-plane Pt lattice
constant.  The distance between interface Co and Pt layers is taken
from Ref.~\onlinecite{wu}, and the distance between Co planes was
chosen to match the bulk Co density.  Our qualitative conclusions do
not depend sensitively on these choices, as we discuss in Sec. III.  We
present results for 1, 2 and 3 monolayers (ML) of Co on 8 layers of
Pt, stacked along the [111] direction.  We generally take the Co layer
stacking to be fcc.  We also study systems with an intermixed
interface (see Fig.~\ref{fig:coords}b).  Computational limitations
preclude the use of a larger super cells; however these smaller unit
cells reduce the overall symmetry of the system, removing some
artifacts present only for ideal systems.
	
We use the local spin density approximation,\cite{perdew} with full
non-collinear spin, and the spin-orbit coupling is included using the on-site
approximation as described in Ref. \onlinecite{sanvito}.  We include
$1.1~{\rm nm}$ of vacuum along the $\hat{\bf z}$ direction and use a
minimal localized atomic orbital basis set (a single $s,p,d$ basis set
for each atom) and norm-conserving pseudopotentials.  Adequately
converging the ground state energy requires a 2d k-point
mesh with mesh spacing $dk=0.18~{\rm nm}^{-1}$.

Using the ground state potential, we find the eigenstates at the
Fermi energy using an approach described in
Refs.~\onlinecite{taylor1,taylor2}.  We evaluate the charge
current and spin torque\cite{nunez} from a Boltzmann distribution of the states
(within the relaxation time approximation):
\begin{eqnarray}
I &=& \frac{e^2\tau}{\hbar\left(2\pi\right)^2} \int d{\bf k}_{\parallel} \left({v}_{\bf k_{\parallel}}^R-{v}_{\bf k_{\parallel}}^L\right) \cdot {E_x} \\
{\bf\Gamma}&=&\frac{e\tau}{\hbar\left(2\pi\right)^2}\int d{\bf k}_{\parallel}
\left(\left(\bf s \times \bf \Delta\right)^R_{\bf k_{\parallel}} -
  \left(\bf s \times \bf \Delta\right)^L_{\bf k_{\parallel}}\right)
{ E_x}
\label{eq:boltzmann}
\end{eqnarray}
where $\tau$ is the lifetime, ${\bf k}_\parallel$ is the 2-dimensional
Bloch
wave vector normal to the current direction (so that the resulting
integral is 1-dimensional).  ${\bf s}$ is the spin operator, and
$\Delta$ is the spin-dependent exchange-correlation potential. The
superscript $R$
($L$) refers to states with positive (negative) group velocity in the
$\hat{\bf x}$-direction.  The resulting torque per current is independent
of scattering time $\tau$.  Converging the 1-dimensional transport integrals requires a
finer mesh spacing ($dk=0.044~{\rm nm}^{-1}$) than needed for the
ground state energy.  While the torque varies strongly with the
magnetization direction, the effective field only varies weakly.  We
present our results in terms of a scaled effective field
\begin{equation}
  \label{eq:scale}
  H_R/J=\frac{\bf\Gamma \cdot [{\bf M}\times(\hat{\bf j}\times\hat{\bf z})]} {|{\bf M}\times(\hat{\bf j}\times\hat{\bf z})|^2} \frac{d}{M_{\rm s} I}
\end{equation}
%\frac{|\bm\Gamma|} {|\hat{\bf M}\times(\hat{\bf j}\times\hat{\bf z})|}
%  \frac{d}{M_{\rm s} I} ,
where $M_{\rm s}$ is the computed magnetization density and $d$ is the
slab thickness.  The result has units of field per bulk current
density, a quantity commonly used to report experimental results.

Eq.~\ref{eq:boltzmann} captures some - but not all - contributions to
the current-induced torque.  In the language of
Ref.~\onlinecite{Freimuth}, we only include the odd torques, and hence
neglect any intrinsic spin Hall effect in the Pt layers or at the
interface.  We also neglect effects from interband scattering on the torques\cite{jungwirth} as well as higher order corrections due to momentum scattering that were considered in Refs. ~\onlinecite{macdonald,wang} for a Rashba model.
Ref.~\onlinecite{Freimuth} shows that for Co/Pt, the
even torque is dominated by the spin Hall effect and the
interfacial contributions are negligible.  We also neglect skew
scattering that could also give a spin Hall effect.  Thus, we only
capture the current-induced torque related to the Rashba model and
neglect all contributions from the spin Hall effect in the bulk of the
Pt.  Recent models of this system employing a Boltzmann model show
that the Rashba and spin Hall effect torques are largely independent
of each other.\cite{stiles:2013} We use the approach described here to
focus on understanding the contribution to the current-induced torque
from the spin-orbit coupling near the interface.

Our calculations are complementary to those of Ref.~\onlinecite{Freimuth}
Those calculations compute the torque for the magnetization parallel
and antiparallel to the interface normal and consider the even and odd
components.  These are equivalent to the damping-like and field-like
torques.  Their calculations include the spin Hall effect in the Pt
and so describe the damping-like torque that we neglect.  Where the
calculations can be compared, we find similar results.  The reduced
computational demands of our approach enable us to explore a wider
range of systems, including different geometries and different
magnetization orientations.

\section{Results}

\begin{figure}
\includegraphics[width=0.8\columnwidth]{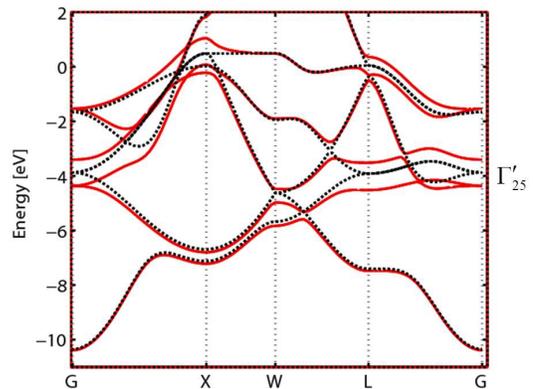}
\caption{(Color online) Band structure of bulk Pt.  Solid lines: with spin-orbit coupling, dotted lines: without spin-orbit coupling.}\label{fig:bands}
\end{figure}

We first consider the important energy scales in the system: the
spin-orbit coupling and the exchange spin splitting.
Fig.~\ref{fig:bands} shows the calculated bulk band structure of
bulk Pt with and without spin-orbit coupling.  The spin-orbit energy
splittings are large at points of high-symmetry: the spin-orbit
splitting at ${\bf k}=0$ of the ${\Gamma}_{25}'$ band is on the order of
$1~{\rm eV}$.  For bulk hcp Co, we find an exchange splitting energy
$\Delta$ of about $1~{\rm eV}$.  We show below that the
exchange splitting in the Co induces exchange splitting (and a small
moment) in the Pt and that the spin-orbit coupling in the Pt induces a
transverse moment in the Co.  Of these proximity effects, the transverse moment on the Co induced by the spin-orbit coupling in the Pt
plays the dominant role in determining the current-induced torques.

\begin{figure}[h!]
\includegraphics[width=1.0\columnwidth]{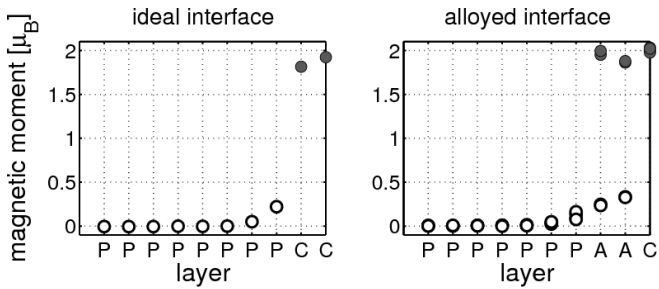}
\caption{Magnetic moment versus layer for (a) 8 layers of Pt (P) - 2
  layers of Co (C), and (b) 8 layers of Pt - 2 alloyed (A) layers - 1
  layer Co, (see Fig.~2b for the alloy coordinates).  The four
  atomic moments in each layer are plotted individually (since the disordered interface has inequivalent atoms in each layer), where filled
  (open) symbols represent Pt (Co). } \label{fig:moments}
\end{figure}

The layer resolved magnetic moments are shown in
Fig.~\ref{fig:moments}a.  The induced moment on the interface Pt layer
varies with the Co coverage, with values of $0.30~\mu_{\rm B},0.22~\mu_{\rm B},~{\rm and}~0.25~\mu_{\rm B}$ for 1 ML, 2 ML, and 3 ML of Co, respectively.
These values are similar to those found in previous
calculations\cite{wu,moulas,lehnert} and to experimental measurements
(Refs.~\onlinecite{ferrer,geissler,valvidares} measure a Pt
interface moment of $\approx 0.2~\mu_{\rm B}$, while
Ref.~\onlinecite{suzuki} measures a moment of $\approx 0.6~\mu_{\rm
  B}$).  Fig.~\ref{fig:moments}b shows the moments in each layer for
the disordered interface geometry showed in Fig.~\ref{fig:coords}b.
The moments on the Pt atoms in the alloyed interface range from
0.25~$\mu_{\rm B}$ to 0.4~$\mu_{\rm B}$.  The importance of this
proximity magnetization in the metal is an open question.  In a recent
experiment, a Au layer was placed between the Pt and the
Co.\cite{parkin}  Au has a lower magnetic susceptibility and a much
smaller induced moment.  It is found that increasing the Au layer thickness reduces the offset in the current-domain wall velocity curves, which is explained by a reduction in the DM interaction.  It's concluded that proximate magnetization plays a central role in one of the important spin-orbit coupling effects at the interface (the DM interaction).\cite{parkin} On the other hand, our calculations suggest that
the transverse spin in the Co induced by the Pt plays a more important
role for the field-like torque, as we discuss at the end of this section.

\begin{figure}[h!]
\includegraphics[width=1.0\columnwidth]{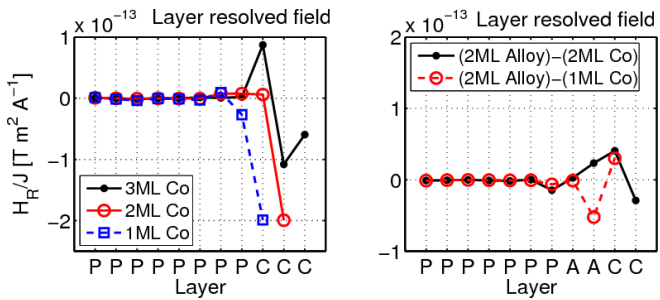}
\caption{(color online) Layer resolved effective field per current density for (a) 1,
  2, and 3 ML of Co with an ideal interface, and (b) 2 alloyed
  interfaces. } \label{fig:torque1}
\end{figure}

Fig.~\ref{fig:torque1}a shows the layer-resolved effective field per current for 1, 2, and
3 ML of Co.  The largest field (or largest torque) is on the Co atoms, although there is
also torque present on the magnetized Pt layer.  The effective field per current for various geometries are
listed in Table~1.  Experimental values range from a high\cite{miron1} of $10^{-12}~{\rm
  T\cdot m^2/A}$ to slightly above\cite{kim0,garello} $10^{-14}~{\rm
  T\cdot m^2/A}$, although the precise value is highly dependent on
system details like layer thicknesses: Ref. \onlinecite{ohno2} finds an order of magnitude difference in the current-induced effective field when the magnetic layer thickness changes from 1 nm to 1.2 nm.  We also find the magnitude depends sensitively on
coverage.  Fig.~\ref{fig:torque1}b shows the effective field for an alloyed interface.
The total magnitude is decreased in all alloyed interfaces
we have investigated, relative to the ideal interface.  The torques are
again predominantly localized on the Co atoms, and they are
nearly oppositely oriented for Co atoms in the alloy layer adjacent to
the pure Pt.  This cancelation is largely responsible for the decrease
in the total current-induced torques.  Note that our result for the 8-3 ML Pt-Co system
is similar to that of Ref. \onlinecite{Freimuth}, despite some slight differences in the
structures used in the two calculations.

The torques are largely unchanged if we use hcp stacking for the Co
layers (see Table 1).  They do change in calculations for larger Co-Pt
and Co-Co interplane distance, but the trends with respect to Co
coverage and alloying are similar.  We highlight the trend that the torques are decreased for imperfect interfaces, and that these decreased values are in semi-quantitative agreement with experiment.

\begin{table}
\begin{tabular}{|c|c|c|c|}
  \hline
  Structure& SO on Pt & $\Delta$ on Pt &  $H_R/J
  \left[10^{-14}~{\rm T\cdot m^2/A}\right]$ \\ \hline
  8-0-1      & $\bullet$ & $\bullet$ & -22.4 \\ \hline
  8-0-2      & $\bullet$ & $\bullet$ &  -17.9 \\ \hline
  8-0-2      & $\bullet$ &   &  -19.5 \\ \hline
  8-0-2      &   & $\bullet$ &  -3.2 \\ \hline
  8-0-2(hcp) & $\bullet$ & $\bullet$ & -18.9  \\ \hline
  8-1-1      & $\bullet$ & $\bullet$ &  -11.1 \\ \hline
  8-2-1 (a)  & $\bullet$ & $\bullet$ &  -3.3  \\ \hline
  8-2-1 (a)  & $\bullet$ &   & -3.9  \\ \hline
  8-2-1 (a)  &   & $\bullet$ & 2.9 \\ \hline
  8-2-1 (b)  & $\bullet$ & $\bullet$ & -3.7 \\ \hline
  8-0-3      & $\bullet$ & $\bullet$ &  -7.7 \\ \hline
  8-2-2      & $\bullet$ & $\bullet$ &  -1.9  \\ \hline
  \hline
\end{tabular}
  \caption{Field-like torque for different film geometries.  The first
  column gives the number of pure Pt layers, the number of alloy
  layers, and the number of pure Co layers.  In the second and third
  columns, no ``$\bullet$" appears if the spin-orbit coupling (second) or
  exchange potential (third) has been set to zero.  The two different
  structures considered for the 8-2-1 geometry are designated (a) and (b).}
  \label{tab:fltorque}
\end{table}

\begin{figure}
\includegraphics[width=.95\columnwidth]{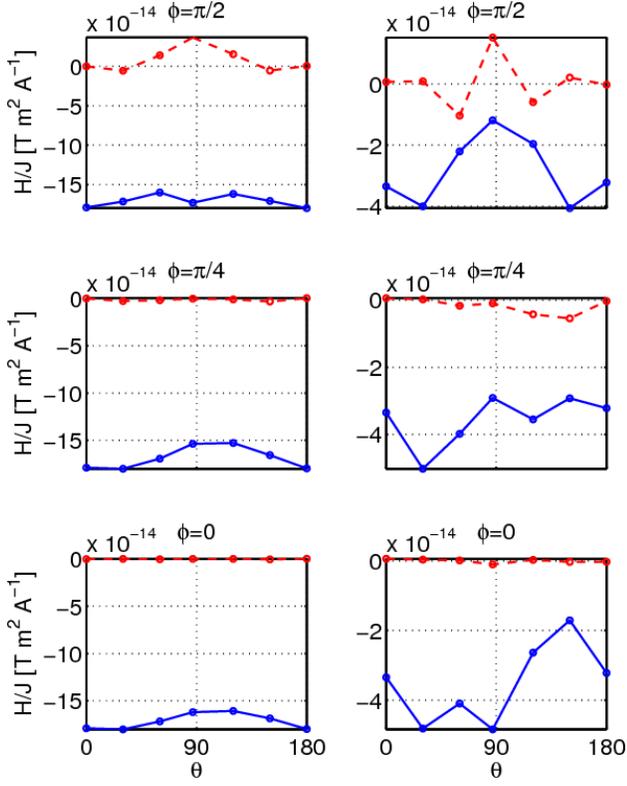}
\caption{(color online) The angular dependence of the effective field (solid blue) and
  ``damping field" (dashed red) magnitude per current density.  The left panel
  is for 2 ML of Co with an ideal interface, while the right panel is
  for 2 layers of alloy+1 layer Co.  The angles are conventional
  spherical coordinates for the coordinate system of
  Fig.~2. } \label{fig:angular}
\end{figure}

As noted above, our calculations include only the odd contribution to the torque,
$\bm\Gamma(\hat{\bf M})=-\bm\Gamma(-\hat{\bf M})$.  This contribution is dominated by the field-like torque, although we also find a nonzero damping-like torque for certain orientations of the magnetization.  %This contribution points mostly along the $\pm \hat {\bf M} \times \left(\hat {\bf j} \times \hat {\bf z}\right)$ direction and is thus dominated by the field-like torque.  However we also find a torque which is orthogonal to $\hat{\bf M} \times \left(\hat {\bf j} \times \hat {\bf z}\right)$ and $\hat {\bf M}$: this is a damping-like torque which is odd in M (the prefactor is odd in magnetization direction).
%to the damping-like torques with a prefactor
%that is odd in the magnetization direction.
Fig.~\ref{fig:angular}a shows the angular dependence of the
current-induced fields responsible for field-like and damping-like
torques.  The left panels are for an ideal 8-2 ML Co-Pt system, and
the right panels are for a disordered interface as shown in
Fig.~\ref{fig:coords}b.  For the Rashba model in the limit where the
exchange potential is much larger than the spin-orbit coupling energy,
the equivalent fields are independent of the magnetic
orientation.\cite{manchon} However, when the exchange potential and
spin-orbit coupling are similar in magnitude, the angular dependence
of the fields found from the Rashba model is similar to that shown here.\cite{Lee2}  Based on Fig.~\ref{fig:angular}, we conclude that the
simple Rashba model description of bilayers accounts reasonably well
for many properties of the torque (at least for an ideal
interface).

%BETTER ARGUMENT HERE ABOUT WHY in-plane symmetry breaking is needed.
We next comment on the damping-like torque calculated in our system.  We first note that if a system is isotropic in the $x-y$ plane, the odd torque is entirely field-like.  This is because the eigenstate have spin components $s_x,~s_z$ components which are even functions of $k_x$, and $s_y$ which is an odd function of $k_x$.\cite{footnote2}  Forming a current-carrying distribution in the $\hat x$-direction leads to a spin
accumulation purely in the $\hat y$-direction, yielding a field like torque.  The inequivalence of $+\hat x$ and $-\hat x$ directions of our lattice implies that the spins have no such symmetries when the magnetization has a $\hat y$ component, so that a spin accumulation of any direction is allowed by symmetry.\cite{footnote3}  This in turn leads to both field-like and damping-like torques.  For real systems with disordered interfaces, we expect the in-plane direction to be relatively isotropic, so that the odd torque is primarily field-like (at least within the relaxation time approximation).

\begin{figure}
\includegraphics[width=0.9\columnwidth]{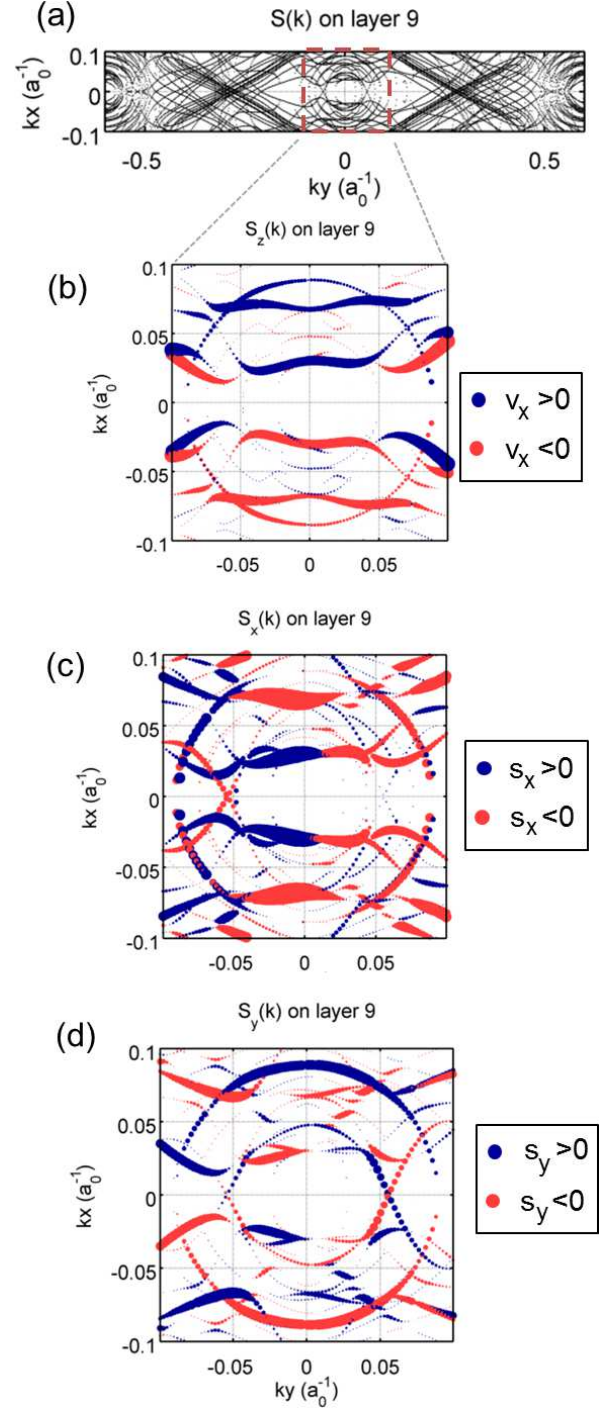}
\caption{(color online) (a) Plot of states at the Fermi level versus wave vector. (b)
  Zoom in of states near the zone center.  The dark blue (light red)
  color indicates states with positive (negative) group velocity in
  the $x$-direction.  The dot size of each state is proportional to
  the magnitude of the state's $z$ component of spin on the interface
  Co layer. (c) The $x$ component of spin on the interface Co layer.
  The magnitude of each state's spin is proportional to the dot
  size, and the colors (shading) indicate the sign of $s_x$.  (d) The
  same plot for the $y$ component of each state's
  spin.  The torque on the interface Co layer
  contributed by each state is proportional to the in-plane spin
  component for that state.}\label{fig:Sk}
\end{figure}

To further illustrate the similarities and differences of the system
with simple Rashba model, we plot the states at the Fermi energy in Fig. \ref{fig:Sk}.
Despite the enormous complexity of the electronic structure, there are similarities
to the Rashba model.  This is shown in Figs. \ref{fig:Sk}c-d, which depict the spin structure
of states near ${\bf k}$=0.  The symmetry of the Fermi surface and ${\bf k}$-dependence of the spin follows from the system symmetry.\cite{footnote3}  The current-induced torque arises when summing over a current carrying distribution of these states (recall the current to be in the $\hat x$ direction).  The sizeable transverse moments of the states, shown in panels (c) and (d) of Fig. \ref{fig:Sk}, are the result of the interaction of the Co orbitals with the spin-orbit potential localized on the Pt.

Summing over states leads to
significant cancellation of the torques.  We find that the average
absolute value of torque from each state is about 10 to 100 times greater
than the integrated total, depending on the specific system.  As
described in Ref. \onlinecite{park1}, this cancellation can be understood
qualitatively in a tight-binding model, where different bands have
different signs of orbital chirality $\pm\left({\bf L}\times{\bf
    k}\right)\cdot\hat{\bf  z}$.\cite{kim:2011}  The addition of spin-orbit
coupling ${\bf L}\cdot{\bf S}$ then (roughly speaking) leads to
alignment of the spin in the $\pm\left({\bf S}\times{\bf
    k}\right)\cdot\hat{\bf z}$ direction, resulting in different signs of
an effective Rashba parameter for different states.\cite{park1}

%\begin{figure}
%\includegraphics[width=0.8\columnwidth]{Sk_FORPAPER.eps}
%\caption{The expectation value of the in-plane spin for states of the 8 layer Pt - 2 layer Co system, near the Brouillon zone center.  The red (blue) arrows indicate states %with positive (negative) $v_z$.} \label{fig:Sk}
%\end{figure}
	
We next attempt to identify the primary source of the torque present
at the interface.  A current-induced torque from a Rashba-like model requires the simultaneous
presence of exchange splitting and spin-orbit coupling.  It's not {\it
  a priori} obvious if the induced exchange present in the Pt is more
or less important than the induced spin-orbit coupling in the Co.  We
first make a general remark about this distinction.

\begin{figure}
\includegraphics[width=0.8\columnwidth]{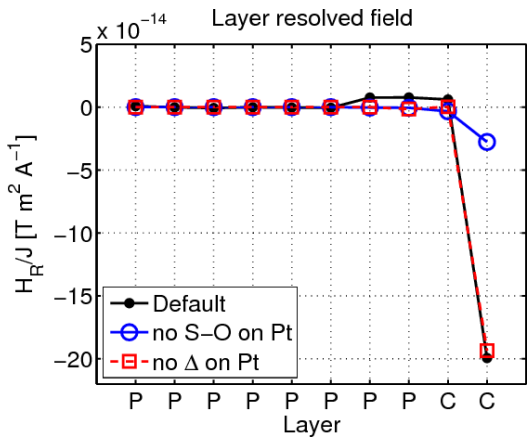}
\caption{(color online) Layer resolved torque for 8 layer Pt - 2 layer Co system.  ``Default" refers to the system with full exchange and spin-orbit coupling.  The other torques are calculated by removing the specified potential from the Pt atoms, as described in the text.  } \label{fig:scenarios}
\end{figure}

The magnetic proximity effect is one of the most dramatic examples of the effect of hybridization between neighboring materials' orbitals.  However, as discussed in Ref. \onlinecite{park1}, there is a similar energy splitting on the Co orbitals due to the interaction with the spin-orbit coupling in the Pt.  This effect is not as obvious as the magnetic proximity effect for two reasons:  the Co levels with which Pt hybridize are pure spin states, which leads to appreciable spin splitting for all Pt states, and a macroscopic magnetization in the Pt interface layer.  On the other hand,  the eigenstates of the atomic spin-orbit potential ${\bf L}\cdot {\bf S}$ are those of the total angular momentum operator ${\bf J}={\bf L}+{\bf S}$.  ${\bf J}$ is not a good quantum number in the symmetry broken crystal field, so the Pt states with which Co hybridize generally do not directly reflect the spin orbit potential.  This obscures the effect of the Pt spin-orbit on the levels in the Co.  Despite being less ``obvious" in this way, the proximate effect of the spin-orbit coupling is quite important, as we show next.

To determine the role of magnetic proximity effect, we remove the exchange splitting on the Pt atoms from the ground state Hamiltonian, and the resulting current-induced torques are calculated as described in Sec. II.  To determine the role of the spin-orbit coupling proximity effect, we remove the spin-orbit potential from the Pt atoms, perform a new ground state calculations, and calculated the current-induced torques.  (We find that removing the Pt spin-orbit potential from the initial ground state and calculating the current-induced torques {\it without} performing a new ground state calculations yields very similar results.)

\begin{figure}
\includegraphics[width=1.\columnwidth]{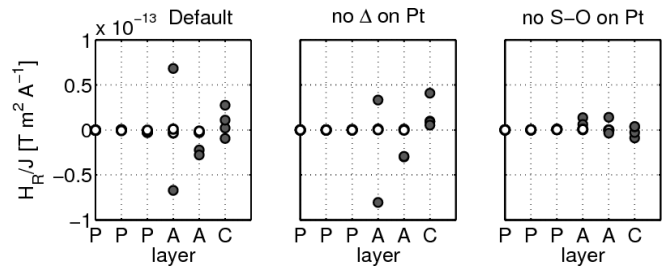}
\caption{
Similar plot as Fig.~7, but with an alloyed interface.  The individual torques on the 4 atoms of each layer are shown (the torques from the bottom 5 Pt layers are omitted from the plot, as they vanish).  Open (filled) circles represent Pt (Co).   } \label{fig:scenariosmix}
\end{figure}

Fig.~\ref{fig:scenarios} shows the results for the ideal interface.  The current-induced effective field is nearly unchanged when the Pt magnetization is removed, and greatly diminished when the Pt spin-orbit coupling is removed.  We conclude that the spin-orbit in the Pt is the main agent behind the total torque.  We additionally repeat this calculation for an alloyed system, with the layer and atom-resolved torques shown in Fig. \ref{fig:scenariosmix}.  A similar scenario holds in this case.  (Note however that the magnitude of the {\it total} torques are not so different in the alloyed case (see Table 1) when we remove the spin-orbit from the Pt; this is due to the significant cancelations that occur when adding the contributions from all the states in the default system.)  It's instructive to evaluate the average absolute value of the torque from all states for these different scenarios.  We find the torques are nearly unchanged for no magnetization on Pt, while they're about 5 times smaller when the spin-orbit is removed from the Pt.  To rationalize this result, we note that the exchange splitting in the Pt is 10 times smaller than in the Co.  On the other hand, the transverse spin density is 3x smaller in the Co than in the Pt.  These properties of the states indicate that the Pt spin-orbit potential is the primary source of the overall torque.

\section{Conclusion}

In summary, we performed first principles calculations of the field-like current-induced torque on a series of Co-Pt bilayers, and from this we conclude that: 1. the torque is very sensitive to system details, such as Co coverage.  This is consistent with experimental data, and the general magnitude of the calculated effective fields is similar to the experimental values.  2.  The angular dependence of the torque is very similar to that predicted by the simple 2-d Rashba model for an ideal interface, but more complicated for alloyed interfaces.  3.  The primary source of the torque is derived from the spin-orbit coupling localized on the Pt interface atoms, which affects the nonequilibrium spin density in the Co interface layer, and drives the current-induced torques.  The extent to which the trends revealed by the first principles calculations can be made more revealing in simpler models is a question for future work.  In addition, this approach may be used to screen different materials combinations in order to anticipate which materials may show the strongest effect.


\begin{thebibliography}{00}
\bibitem{stiles} D. C. Ralph and M. D. Stiles, J. Magn. Magn. Mater. {\bf 320}, 1190 (2008).

\bibitem{katine} J. A. Katine and E. E. Fullerton, J. Magn. Magn. Mater. {\bf 320}, 1217 (2008).

\bibitem{chien}  W.-G. Wang, M. Li, S. Hageman, and C. L. Chien, Nat. Mat. {\bf 11}, 64 (2011).

\bibitem{ohno0} D. Chiba, M. Sawicki, Y. Nishitani, Y. Nakatani, F. Matsukura, and H. Ohno, Nature {\bf 455}, 515 (2008).

\bibitem{miron1}I. M. Miron, G. Gaudin, S. Auffret, B. Rodmacq, A. Schuhl, S. Pizzini, J. Vogel, and P. Gambardella, Nat. Mat. {\bf 9}, 230 (2010).

\bibitem{miron2} I. M. Miron, T. Moore, H. Szambolics, L. D. Buda-Prejbeanu, S.
Auffret, B. Rodmacq, S. Pizzini, J. Vogel, M. Bonfim, A. Schuhl,
and G. Gaudin, Nat. Mater. {\bf 10}, 419 (2011).

\bibitem{miron3} I. M. Miron, K. Garello, G. Gaudin, P.-J. Zermatten, M. V.
Costache, S. Auffret, S. Bandiera, B. Rodmacq, A. Schuhl, and
P. Gambardella, Nature (London) {\bf 476}, 189 (2011).

\bibitem{liu1} L. Liu, O. J. Lee, T. J. Gudmundsen, D. C. Ralph, and R. A.
Buhrman, Phys. Rev. Lett. {\bf 109}, 096602 (2012).

\bibitem{ando} K. Ando, S. Takahashi, K. Harii, K. Sasage, J. Ieda, S. Maekawa,
and E. Saitoh, Phys. Rev. Lett. {\bf 101}, 036601 (2008).


\bibitem{kim0} J. Kim, J. Sinha, M. Hayashi, M. Yamanouchi, S. Fukami, T. Suzuki, S. Mitani, and H. Ohno, Nat. Mat. {\bf 12}, 240 (2013).

\bibitem{parkin} K.-S. Ryu, L. Thomas, S.-H. Yang, and S. Parkin, Nat. Nanotech. {\bf 8}, 527 (2013).

\bibitem{beach} S. Emori, U. Bauer, S.-M. Ahn, E. Martinez, and G. S. D. Beach, Nat. Mat. {\bf 12}, 611 (2013).

\bibitem{fan}  X.Fan, j. Wu, Y. P. Chen, M. J. Jerry, H. W. Zhang, J. Q. Xiao,
 Nature Communications, {\bf 4}, 1799 (2013).

\bibitem{ralph} L. Liu, C.-F. Pai, Y. Li, H. W. Tseng, D. C. Ralph, R. A. Buhrman, Science {\bf 336}, 555 (2012).

\bibitem{manchon} A. Manchon and S. Zhang, Phy. Rev. B  {\bf 78}, 212405 (2008).

\bibitem{Dzyaloshinskii} I. E. Dzyaloshinskii, Sov. Phys. JETP \textbf{5}, 1259 (1957).

\bibitem{Moriya} T. Moriya, Phys. Rev. \textbf{120}, 91 (1960).

\bibitem{thiaville} A. Thiaville, S. Rohart, \'{E}. Ju\'{e}, V. Cros, and A. Fert, EPL {\bf 100}, 57002 (2012).

\bibitem{footnote1} The choice of vector components along which to project the torque is arbitrary, but the conventional choice described here is convenient for these systems.  There can be additional angular dependence of the torque component's magnitudes, as we show in Fig. \ref{fig:angular}.

\bibitem{macdonald} D. A. Pesin and A. H. MacDonald, Phys. Rev. B {\bf 86}, 014416 (2012).

\bibitem{wang} X. Wang and A. Manchon, Phys. Rev. Lett. {\bf 108}, 117201 (2012).

\bibitem{duine2012} E. van der Bijl and R. A. Duine, Phys. Rev. B {\bf 86}, 094406 (2012)

\bibitem{Kim2012} K.-W. Kim, S. M. Seo, J. Ryu, K.-J. Lee, and H.-W. Lee, Phys. Rev. B {\bf 85}, 180404 (2012).

\bibitem{Freimuth} F. Freimuth, S. Bl\"ugel, and Y. Mokrousov,  arXiv:1305.4873 (2013).

\bibitem{ferrer} S. Ferrer, J. Alvarez, E. Lundgren, X. Torrelles, P. Fajardo, and F. Boscherini, Phys. Rev. B {\bf 56}, 9848 (1997).

\bibitem{lundgren} E. Lundgren, B. Stanka, M. Schmid, and P. Varga, Phys. Rev. B {\bf 62}, 2843 (2000).

\bibitem{wu} R. Wu, C. Li, and A. J. Freeman, J. Magn. Magn. Mat. {\bf 99}, 71 (1991).

\bibitem{perdew} J. P. Perdew and A. Zunger, Phys. Rev. B {\bf 23} 5048 (1981).

\bibitem{sanvito}L Fern\'{a}ndez-Seivane, M. A. Oliveira, S. Sanvito, and J. Ferrer, J. Phys. Cond. Mat. {\bf 18}, 7999 (2006).

\bibitem{taylor1} J. Taylor, H. Guo, and J. Wang, Phys. Rev. B {\bf 63}, 245407 (2001).

\bibitem{taylor2} J. Taylor, PhD Thesis, McGill University (2000).

\bibitem{nunez} A. N\'{u}\~{n}ez and A. H. MacDonald, Sol. State Comm. {\bf 139}, 31 (2006).

\bibitem{jungwirth} H. Kurebayashi, Jairo Sinova, D. Fang, A. C. Irvine, J. Wunderlich, V. Nov\`{a}k, R. P. Campion, B. L. Gallagher, E. K. Vehstedt, L. P. Zarbo, K. V\`{y}born\`{y}, A. J. Ferguson, and T. Jungwirth, arXiv:1306.1893 (2013).

\bibitem{stiles:2013} P. M. Haney, H.-W. Lee, K.-J. Lee, A. Manchon, and M. D. Stiles, Phys. Rev. B {\bf 87}, 174411 (2013).

\bibitem{moulas} G. Moulas, A. Lehnert, S. Rusponi, J. Zabloudil, C. Etz, S. Ouazi, M. Etzkorn, P. Bencok, P. Gambardella, P. Weinberger, and H. Brune, Phys. Rev. B {\bf 78}, 214424 (2008).

\bibitem{lehnert}    A. Lehnert, S. Dennler, P. B{\l}o\~{n}ski, S. Rusponi, M. Etzkorn, G. Moulas, P. Bencok,
P. Gambardella, H. Brune, and J. Hafner, Phys. Rev. B {\bf 82}, 094409 (2010).

\bibitem{geissler} J. Geissler, E. Goering, M. Justen, F. Weigand, G. Sch\"{u}tz, J. Langer, D. Schmitz, H. Maletta, and R. Mattheis, Phys. Rev. B {\bf 65}, 020405(R) (2001).

\bibitem{valvidares} S. M. Valvidares, T. Schroeder, O. Robach, C. Quir\'{o}s, T.-L. Lee, and S. Ferrer, Phys. Rev. B {\bf 70}, 224413 (2004).

\bibitem {suzuki} M. Suzuki, H. Muraoka, Y. Inaba, H. Miyagawa, N. Kawamura, T. Shimatsu,
H. Maruyama, N. Ishimatsu, Y. Isohama, and Y. Sonobe, Phys. Rev. B {\bf 72}, 054430 (2005).

\bibitem{garello}K. Garello, I. M. Miron, C. O. Avci, F. Freimuth, Y. Mokrousov, S. Bl\"{u}gel, S. Auffret, O. Boulle, G.Gaudin, and
P. Gambardella, Nat. Nanotech. {\bf 8}, 587 (2013).

\bibitem{ohno2} T. Suzuki, S. Fukami, N. Ishiwata, M. Yamanouchi, S. Ikeda, N. Kasai, and H. Ohno, Appl. Phys. Lett. {\bf 98},
142505 (2011).

\bibitem{Lee2} K.-J. Lee (private communication).

\bibitem{footnote2}  The relevant symmetry operation is reflection in the $x-y$ plane + time reversal.

\bibitem{footnote3}  If the magnetization is in the $x-z$ plane, the system is invariant under operations of reflection about the $x-z$ plane (taking $y\rightarrow -y$) followed by time reversal.  This implies degenerate states at $\left(k_x,k_y\right)$ and $\left(-k_x,k_y\right)$, with equal $s_x$ and $s_z$, and opposite $s_y$, yielding exclusively a field-like torque.  A similar argument can not be made for magnetization in the $y-z$ plane, because the lattice is not invariant under $x\rightarrow -x$.

\bibitem{park1} J.-H. Park, C. H. Kim, H.-W. Lee, J. H. Han, Phys. Rev. B {\bf 87}, 041301(R) (2013).

\bibitem{kim:2011} S. R. Park, C. H. Kim, J. Yu, J. H. Han, and C. Kim, Phys. Rev. Lett. {\bf107}, 156803 (2011).


\end{thebibliography}
\end{document}